\documentclass[prb,twocolumn]{revtex4-1}
\usepackage{amsmath}
\usepackage{latexsym}
\usepackage{amssymb}
\usepackage{mathrsfs}
\usepackage{bm}
\usepackage{indentfirst}
\usepackage{natbib}
\usepackage{fancyhdr}
\usepackage{graphicx}
\usepackage{relsize}
\usepackage[a4paper, margin=1.6cm,top=3.5cm, bottom=2.5cm, includefoot]{geometry}
\usepackage{epstopdf}
\usepackage{comment}
\usepackage{wrapfig}
\usepackage{placeins}
\usepackage{float}
\usepackage{sidecap}
\usepackage{lmodern}
\usepackage[T1]{fontenc}
\usepackage{verbatim}
\usepackage{enumerate}
\usepackage{setspace}
\usepackage{color}
\usepackage{url}

\def\onedot{$\mathsurround0pt\ldotp$}
\def\cdddot#1{
  \mathbin{\vcenter{\baselineskip.67ex
    \hbox{\onedot}\hbox{\onedot}\hbox{\onedot}%
  }}%
}
\def\fdot#1{
  \mathbin{\vcenter{\baselineskip.6ex
    \hbox{\onedot}\hbox{\onedot}\hbox{\onedot}\hbox{\onedot}%
  }}%
}
\usepackage{appendix}

\begin{document}

 \title{Multi-body quenched disordered $XY$ and $p$-clock models on random graphs}

\author{Alessia Marruzzo$^{1,2}$, Luca Leuzzi$^{2,1}$}

\affiliation{$^1$ Dipartimento di Fisica, Universit\`a {\em Sapienza},
  Piazzale Aldo Moro 5, I-00185, Rome, Italy\\ $^2$ Soft and Living
  Matter Lab., Rome Unit of CNR-NANONTEC, Institute of Nanotechnology,
  Piazzale Aldo Moro 5, I-00185, Rome, Italy}

\begin{abstract}
The $XY$ model with four-body quenched disordered interactions  and its discrete $p$-clock proxy are
studied on bipartite random graphs by means of the cavity method.  The
phase diagrams are determined from the ordered case to the spin-glass
case.  Dynamic, spinodal and thermodynamic transition lines are
identified by analyzing free energy, complexity and tree
reconstruction functions as temperature and disorder are changed.  The study of the
convergence of the $p$-clock model to the $XY$ model is performed down
to temperature low enough to determine all relevant transition points
for different node connectivity.
\end{abstract}

\maketitle

\section{Introduction}

The aim of this paper is to provide a general theoretical framework for the prediction of specific transitions from incoherent to coherent regimes of multi-body  interacting wave modes  in random networks with quenched random valued interaction couplings. 
The nature of an highly-correlated phase  at low temperatures specifically depends  on the disordered network of interactions and on the strength of the nonlinear interaction couplings.  When considered non-perturbatively, quenched disorder, i.e. , the disorder that does not change on the time-scales of the waves dynamics, can yield a glassy behavior in given systems. This includes the possibility of displaying a large number of degenerate state realizations, so large to yield an extensive configurational entropy.\\
 In particular, we have in mind random lasers, in which the feedback is provided by the multiple scattering of light inside the optically active random media providing gain\cite{Cao03, Skipetrov03, vanderMolen06a, Tureci08, Wiersma08, Leuzzi09a, Zaitsev10, ElDardiry10, Andreasen11, Antenucci15a}. 
  In these situations, modes may in general exhibit complicated spatial extensions and 
   their interaction strengths, related to the spatial superposition of the electromagnetic fields modulated by a heterogeneous nonlinear optical susceptibility,  are disordered. These disordered couplings can induce strong {\em frustration} \footnote{For  the definition of frustration in disordered systems, introduced by Anderson \cite{Anderson77}, see, e.g., Refs. \cite{Anderson78,Mezard87,Leuzzi07b} and references therein.}, that is: a  given set of interacting modes with a given interaction network is not able to find a single optimal configuration but only many energetically (or entropically) competing sub-optimal ones.
 Representing mode phases, $\varphi(t)$, by means of continuous planar XY-like spins, $\vec{\sigma}(t) = \left(\cos\varphi(t),\sin\varphi(t)\right)$, and applying a statistical mechanics approach we can identify different thermodynamic phases.\\  
 \indent 
 XY models with linearly interacting spins are well known systems in statistical mechanics, displaying important physical insights and applications, starting from the  Berezinsky-Kosterlitz-Thouless transition in two dimensions \cite{Kosterlitz72} and moving to, e.g., the transition of liquid helium to its superfluid state \cite{Brezin82, Brezin10}, the roughening transition of the interface of a crystal in equilibrium with its vapor \cite{Cardy96}, or synchronization problems related to the Kuramoto model \cite{Kuramoto75,Acebron05,Gupta14,PhysRevE.72.066127}. 
 Statistical inference on the XY model has been, as well, recently analysed\cite{Tyagi15}: the so-called inverse XY problem deals with the reconstruction of the network of interactions and with the inferring of the interaction couplings from spin configurations.\\
  As mentioned above, our present motivations to study multi-body XY models are to be mainly found in optics, to describe, e.g., the nonlinear interaction among electromagnetic modes\cite{Conti11,Antenucci15a,Marruzzo15,Antenucci15e,Nixon13}.\\
\indent 
A so-called random laser is characterized by a multi scattering medium and a gain material, which may or may not coincide\cite{PhysRevLett.82.2278,Folli12}.  
Lasing action is induced by an external source pumping the gain material to population inversion. 
When  enough energy is pumped into the system, i. e., when the system is {\em above threshold}, lasing is triggered. 
At difference with  ordinary lasers, where the lasing modes
 are usually the longitudinal Gaussian modes, 
 in random lasers the spatial distributions of the modes is in general more 
  complex and difficult to predict related to the multiple scattering of light caused by the disorder in the linear refractive index. 
 Moreover, the mode frequencies, related to the frequencies of the cold-cavity modes\cite{Sebbah02,Andreasen11} and to the gain curve, are not in general equispaced\cite{Cao05}. 
  We know that the strength of the non-linear interaction is related to the spatial overlap of the modes while the frequencies of the non-linearly interacting modes satisfy the Frequency-Matching condition\cite{SargentIII78}(see Sec. \ref{sec:model_description}). Then, the characteristics of random lasers can yield a random network of mode-couplings.
  To investigate the effect of couplings dilution and their randomness, we adopt a diluted random graphs description. This  can be, as well,  applied to the study of interference effects among neighborhood modes in light guides \cite{Soldano95} or to model the mode-coupling and the related collective behavior of light in  experimentally designed networks of interacting apart lasers 
\cite{Jian05,Rogister07,Nixon13}.\\ 
In a previous work \cite{Marruzzo15} we analysed the role of coupling dilution in statistical mechanical models for lasers,
both below and above threshold.
In this work we move to consider, next to random dilution, the role of frustration, induced by the randomness in the {\em value} of the couplings, that is extracted  according to a generic distribution over a dominion including both positive and negative values. Our main predictions are about the system phase diagram and the configurational entropy, here termed complexity function, that is a typical observable for the glassy phase of disordered systems.\\
Our results indicates that, for negligible nonlinearity, i.e., low pumping or high temperature, all the modes will oscillate independently in a continuous wave noisy regime (``paramagnetic-like'' phase). For a strong interaction and a limited fraction of negative nonlinear coefficients, all the modes will oscillate coherently. This can be realized both as a ``ferromagnetic-like'' regime or as an unmagnetized ordered mode-locked regime corresponding to standard ultra-short laser systems\cite{Svelto98,Haus00}. In intermediate regimes, when  the probability of having negative bonds increases, the tendency to oscillate synchronously turns out to be hindered by disorder-induced frustration, resulting into a glassy regime, characterized by means of Replica Symmetry Breaking (RSB) theory. These three regimes are identified by \textcolor{black}{different sets}  of {\em marginals}, that is, of probability distributions of the values of the mode phases:
\begin{itemize}
\item{CW. ---}
The continuous wave regime displays a uniform marginal for all modes. 
\item{ML. ---} 
In the lasing mode-locked regime, the majority of marginals are single-peaked around the same value of the phase as in ferromagnetic-like phases, i.e., all these modes acquire the same phase. When frequencies are added in the description\cite{Marruzzo15}, another mode-locked regime can occur characterized by marginals single peaked on a phase value linearly dependent on the mode frequency.
 \item{GRL. ---}
 In the frustrated disordered  regime a glassy random laser occurs: each mode displays a nontrivial single-peaked (or multi-peaked) marginal but the peaks are at  different phase values for each mode.
 \end{itemize}
 To study the stationary states of this problem, we can use tools known in the theory of spin glasses. 
  Indeed, the main original results of this work concern the application of the 1-Step Replica Symmetry Breaking  Cavity Method (1RSB-CM) to continuous spin models on diluted bipartite graphs. The attention on diluted spin-glass systems has mainly come from the connection between statistical mechanics of disordered systems and hard optimization problems.  This is the case, for instance, of the $K$-satisfiability problem (see for example \cite{PhysRevE.66.056126}), or error correcting codes\cite{0305-4470-36-43-032}, only to name a few. Models with finitely connected spins emerge also in neural networks\cite{derrida_87} and immune networks\cite{barra_10}.\\ 
\indent 
This work is organized as follows: in Sec. \ref{sec:model_description} we derive the model starting from the optics problem  and we discuss its limit of validity; in Sec. \ref{sec:cavity_equations} we present the Cavity Method on diluted factor graphs considering also the scenario in which many metastable states appear in the phase space. In this section we introduce the $p$-clock model used in numerical algorithms. In Sec. \ref{sec:res} the results obtained are shown and discussed; Sec. \ref{sec:end} is reserved to a summary of the results and to discussions. In App. \ref{app:rsb_m_1} we report more details concerning the derivation of the analytic expressions while in App. \ref{app:pclock} we report the details of the numerical implementations.

\section{The model}
\label{sec:model_description}
The leading model for nonlinearly
interacting waves, under  stationarity conditions,  consists of a system of Langevin equations with a generalized temperature, whose role is to take into account random noise forces coming from the interaction with the outside bath and from spontaneous emission\cite{Viviescas03,Antenucci15e}:
\begin{equation}\label{eq:langevin}
\dot{a}_n = - \frac{\partial \mathcal{H}}{\partial a_n^*} + \eta_n(t)
\end{equation}
where $a_n(t)$ indicates the complex amplitude of mode $n$, i.e., we are assuming that at the steady state the field can be expanded as:
\begin{equation}\label{eq:expansion_e}
\bm{E}(\bm{r},t) = \sum_n a_n(t) e^{- i \omega_n t}\bm{E}_n(\bm{r}) + c.c.  
\end{equation}
with $a_n(t)$ varying on longer time scales than $1/\omega_n$.
$\eta_n(t)$ in Eq. (\ref{eq:langevin}) is a white noise for which:
\begin{equation}\label{eq:noise}
\langle\eta_n(t) \eta_{n'}(t')\rangle = 2 T \delta_{n n'}(t-t')
\end{equation}
$T$ gives the strength of noise, which, in the stationary regime, is independent of time and it can be described as an effective temperature. 
The Hamiltonian is given by:
\begin{eqnarray}
\label{eq:H}
\mathcal{H}&=&-\sum_{jk}^{1,N} g_{jk} a_j^*a_k 
-\sum^{}_{\bm k_4} \tilde{J}_{\bm k_4} a^*_{k_1}a_{k_2} a_{k_3}^*a_{k_4} 
\\
\nonumber
&&\hspace*{5.4cm}+\quad \mbox{c.c.}
\end{eqnarray}
where $\{\bm k_4\}=\{k_1,k_2,k_3,k_4\}$ and the interaction is among the quadruplets that satisfy the Frequency Matching Condition, i.e.,
$$
|\omega_{k_1}-\omega_{k_2}+\omega_{k_3}-\omega_{k_4}| \lesssim \delta \omega
$$ 
where $\omega_{k_i}$ is the frequency of mode $i$, $\delta \omega$ is the linewidth; the $g_{i j}$ are non-zero only for frequency overlapping modes.\\  
 We recall that the two terms of the Hamiltonian (\ref{eq:H}) come from a perturbation expansion, up to third-order, around the non-lasing solution of the Maxwell-Bloch equations describing the dynamics of the electric field coupled to two-level atoms\cite{SargentIII78,Hackenbroich05}. In this approximation, the master equation describing the dynamics of the complex amplitude $a_n(t)$ does not take into account gain saturation, i.e., the fact that in lasers the gain depends on the intensity of the modes, decreasing 
as the modes intensify. 
Indeed, in closed cavity lasers\cite{Haus00,Gordon03a,GorFis03b_fromme},  the gain is usually defined through:
\begin{equation}
g = \frac{g_0}{1+\mathcal{E}/E_{sat}}
\end{equation}
 where
$\mathcal{E}$ is the total optical intensity: $\mathcal{E} \propto \sum_n a^*_n a_n$ that is related to the total optical power pumped into the system; $E_{sat}$ is the saturation power of the amplifier.
Following Gordon and Fisher\cite{Gordon02}, we can introduce a simpler model for  gain saturation that, however, preserves the stability and stationarity of the laser 
 and simplifies the theory: 
 at any instant $g$ is assumed
to take the value needed for maintaining $\mathcal{E}$ constant. As $\mathcal{E}$ is larger than some given threshold (to be determined self-consistently as a critical point of the theory) it  yields the atomic population inversion necessary to have amplified stimulated
emission.\\
The model variables are then $N$ complex numbers $a_n$ 
 satisfying the global
constraint on their magnitudes in the $N$-dimensional space{\cite{Gordon02,Antenucci15e}}, 
$\mathcal{E}=\sum_k |a_k|^2=\mbox{constant}$, at an effective canonical equilibrium.\\ 
Expansion Eq. \eqref{eq:expansion_e} is not unique and a linear transformation can be used in order to obtain a diagonal linear interactions\cite{eremeev11}. 
Writing then the complex amplitudes, $a_k(t)$s, through their amplitudes and phases, $a_k(t)=A_k(t) e^{i \varphi(t)}$, Eq. \eqref{eq:H} becomes:
\begin{eqnarray}
{\cal H}&=&-\sum_{k}^{1,N} g_{kk} A_k^2 
-\sum^{}_{\bm k_4} A_{k_1} A_{k_2} A_{k_3} A_{k_4}  
\nonumber
\\
&& \times \Bigl[ \tilde{J}^{(R)}_{\bm k_4}  \cos\left(\varphi_{k_1}-\varphi_{k_2} +\varphi_{k_3}-\varphi_{k_4}\right)\Bigr.\nonumber \\
&&\Bigl.+ \tilde{J}^{(I)}_{\bm k_4} \sin\left(\varphi_{k_1}-\varphi_{k_2} +\varphi_{k_3}-\varphi_{k_4}\right)\Bigr] 
\label{eq:H2}
\end{eqnarray}
where $\tilde{J}=\tilde{J}^{(R)} + \imath \tilde{J}^{(I)}$.\\
 Under the so-called \emph{free-running approximation}\cite{Antenucci15e}, according to which the phases of the lasing modes are assumed to be uncorrelated and the modes oscillate independently from each other, the non-linear interaction concerns the intensity alone. This case, analysed previously\cite{PhysRevA.81.023822,matrix_sc_all}, neglects  any phase-locking term that could drive the system into the mode-locking regime\cite{Haus00}. Within our approach we can include this term in the description. Moreover, concerning random lasers, it has been seen that, increasing the optical pumping in presence of frustration, the lasing threshold occurs after phase-locking\cite{Antenucci15a,Antenucci15b,Antenucci15c,Antenucci15d}.\\
Being, then, primarily interested in the dynamics of the mode phases, we will work within the \emph{quenched amplitude approximation}: we can consider observing the
system dynamics at a time-scale longer than the one of the phases but
sensitively shorter than the one of the magnitudes, thus regarding the
$A_k$s as constants.
 In this case the first linear term in Eq. \eqref{eq:H} is an irrelevant constant in the dynamics\cite{Antenucci15e,Gordon02}. Rescaling then the non-linear couplings with the mode amplitudes, i.e., $\tilde{J} A_{k_1} A_{k_2} A_{k_3} A_{k_4} \equiv J$,  we obtain the Hamiltonian of the $XY$ model with  four-body interaction terms\footnote{The phases become the relevant variables also when there is energy equipartition, that is $A_k \sim A \forall k$.} i.e., 
 \begin{eqnarray}
 {\cal H}=-\sum^{}_{\bm k_4} \big[ & & J^{(R)}_{\bm k_4}  \cos\left(\varphi_{k_1}-\varphi_{k_2} +\varphi_{k_3}-\varphi_{k_4}\right)\\
 \nonumber
 & &  + J^{(I)}_{\bm k_4} \sin\left(\varphi_{k_1}-\varphi_{k_2} +\varphi_{k_3}-\varphi_{k_4}\right) \big] 
 \label{eq:H_third}
 \end{eqnarray}     
\indent
 We note that, for the case of zero stochastic noise, that is $T\rightarrow 0$ in Eq. \eqref{eq:langevin},  Eq. \eqref{eq:H} takes the form of a nonlinear Schr\"{o}dinger equation (NLSE) if the free-running approximation is considered, i.e., $\sum^{}_{\bm k_4} a^*_{k_1}a_{k_2} a_{k_3}^*a_{k_4} \rightarrow\left(\sum_{k_1,k_2}|a_{k_1}|^2|a_{k_2}|^2\right)$. 
  The disordered NLSE in one dimension is one of the simplest systems where  the interplay between localization, caused by a disordered background, and non-linearity can be studied (for a review see \cite{Basko20111577}  and references therein). The main question is whether the nonlinearity is able to destroy localization eventually leading to equilibration. This effect is related to very interesting physical phenomena, e.g., spreading of a localized wave packet or thermalization\cite{Mulansky09,Kottos11}. An active area of research is also the behaviour of cold atoms and Bose-Einstein Condensate in presence of both, a random potential and non-linearity: the dynamics of the disordered Bose-Einstein Condensate, following the Gross-Pitaevskii equation at zero $T$, describes  the phenomenon of the insulator-superfluid transition\cite{shapiro_12}.\\  
  In this work, instead, we want to focus on the effects of the quenched disorder in the non-linear interaction couplings. The possible disorder in the linear couplings plays here a minor role. Indeed, it has been shown that the behaviour of the system in the stationary lasing regime is independent on the distribution of the $g_{kk}$s\cite{Antenucci15c} that can then be considered as independent of the frequency.\\  
 \indent 
 Concerning the non-linear couplings, we note that the $\tilde{J}_{\bm k_4}$s depend on the mode spatial extensions and
  on the nonlinear optical response tensor $\hat\chi^{(3)}$ of
the  random medium:
\begin{eqnarray}
J_{\bm k_4}&\propto & \int_V d\mathbf r~ \hat\chi^{(3)}(\mathbf r; \nu_{k_1},\nu_{k_2}, \nu_{k_3},\nu_{k_4})
\\
\nonumber
&&\qquad \qquad \qquad  \fdot \quad   ~
{\mathbf E}_{k_1}(\mathbf r )  \mathbf{E}_{k_2}(\mathbf r)  \mathbf{E}_{k_3}(\mathbf r)  \mathbf{E}_{k_4}(\mathbf r)
    \end{eqnarray}
Because of the partial knowledge of the mode spatial distribution and the very
poor knowledge of the nonlinear response so far in random media, both theoretically and in experiments, the
distribution describing the values  of $J$'s can be taken from any physically
reasonable arbitrary probability distribution. If the lasing
transition occurs as a statistical mechanical phase transition its behavior will be universal and independent from
the details of the coupling's distribution.  We will take, e. g., the
bimodal distribution
 \begin{equation}\label{eq:rho}
 P(J_{\bm k_4}) = (1-\rho) \delta(J_{\bm k_4}+J) + \rho ~\delta(J_{\bm k_4}-J)
 \end{equation}
 where $J$ is the material dependent mean square displacement of the couplings distribution and
 $\rho\in [0.5:1]$ is the probability of having ferromagnetic-like
 couplings.
     As external driving parameter we will use the dimensionless quantity $\beta J$, that, in the photonic system, is related to the pumping intensity $\mathcal{P} \propto \mathcal{E}/N$:
     $\beta J = \tilde{J} \mathcal{P}^2/T$.
In the following we will restrain to real valued $J$'s for simplicity.


\section{Replica Symmetry Breaking phase transition}
\label{sec:cavity_equations}
We present our study on continuous ($\varphi\in [0:2\pi)$) and discrete ($\varphi=2\pi m/p$, $m=0,\ldots, p-1$) angular spin models with
quenched disordered interaction on bipartite random regular graphs,
i.e., random graphs whose connectivity at each (variable and
functional) node is fixed. In particular, we analyse the
transition to the one-step Replica Symmetry Breaking (1RSB) clustering
of solutions for the marginals at low temperature occurring in presence of
strong, disorder induced, frustration.  In frustrated systems, indeed, the phase space can decompose into many clusters, 
 termed {\em pure
  states} in the statistical mechanical framework\cite{Mezard01}. In the fragmented
space of solutions any two solutions chosen at random can belong
either to the same cluster or to two disjoint clusters.  The
probability of this latter event is equal to the so-called RSB
parameter $x\in[0:1]$. 

 Each  pure state displays a free energy
$F_n$, $n=1,\ldots, \mathcal{N}_{\rm states}$, in terms of which we can define a probability measure over the states $n$:
\begin{equation}\label{eq:weight_sec}
w_n(x)\equiv \frac{e^{-x\beta F_n}}{\Phi(x,\beta)}
\end{equation}
where 
\begin{equation}\label{eq:gen_part}
\Phi(x,\beta) \equiv \sum_n e^{-x\beta F_n}
\end{equation}
is the partition function in the rescaled effective inverse temperature  
$ \beta x$. From Eq. \eqref{eq:weight_sec}, we can see that the effect of $e^{-x \beta F_n}$ is to weight the pure state $n$ among the set of all states as $e^{- \beta H(\{\varphi\})}$ weights the configuration $\{\varphi\}$ in a single state. In the next section, we will discuss a method, known as the 1-Step Replica Symmetry Breaking Cavity Method (1RSB-CM), to compute the properties of the pure states. For example, we will determine the number of states with free-energy $\phi$, $\mathcal{N}(\phi)=e^{N \Sigma(\phi)}$, where $\Sigma(\phi)$ is known as the Complexity. Starting from the Replica Symmetric Cavity Method (RSCM), we will proceed quickly with the derivation of the RSB case, the interested reader can find more details in \cite{Mezard09,Mezard01}.

\subsection{Message passing}
In order to approach the study of multi-body, else said nonlinearly,
interacting systems we make use of bipartite graphs.  A bipartite
graph is made of $N$ variable nodes (labeled by $i,j,\ldots$) and of $M$
functional nodes (labeled by $a,b,\ldots$) representing the
interactions among variables as a node connecting the variables.
We name the message passing from $i$ to $a$ as $\nu_{ia}$, that is the probability distribution of variable node $i$ in a modified graph where the link between $i$ and $a$ has been cut. 
Analogously, $\hat\nu_{ai}$ is the message from a functional node $a$ to a
variable node $i$. $\hat\nu_{ai}$ represents the probability distribution of variable node $i$ in a modified graph where $i$ is linked only to $a$. The set of all messages is termed $\{\underline{\nu},
\underline{\hat{\nu}}\}$. We will indicate with $\partial a$($\partial i$) the set of all neighbouring nodes of $a$(all neighbouring function nodes of $i$).\\
  Within the Replica Symmetric Cavity Method (RSCM)\cite{Mezard01} approach, one assumes that, at the steady state, correlations among variables decrease exponentially with the distance\footnote{In a random factor graph we can define as distance the number of function nodes encountered in the path connecting two variable nodes.}, a property  characteristic of pure states\cite{sg_pedestrians}. Being in the diluted case the length of the loops $\mathcal{O}\left(\log{N}\right)$, one can consider that if a link $i ~ a$ is cut the other variables neighbours of $a$, i.e., $\partial a \setminus i$, become uncorrelated.
The marginals, $\nu_{i a}$ and $\hat{\nu}_{a i}$ can then be written as:
\begin{eqnarray}
\nonumber
\hat{\nu}_{a i}\left(\varphi\right) &=&  
\frac{1}{\hat{z}_{a i}} \int_{0}^{2 \pi} \prod^{l=1,k-1}_{j_l \in \partial a \setminus i} 
d \varphi_{j_l} \nu_{j_l a} \left(\varphi_{j_l} \right) 
\\
\label{eq:uf}
&&\qquad\qquad\times 
\psi(\{ \varphi_{j\in\partial a\setminus i}, \varphi\}|J)\\
\label{eq:update_variab}
\nu_{i a}\left(\varphi\right) &=& 
\frac{1}{z_{i a}} \prod_{b \in \partial i \setminus a} \hat{\nu}_{b i}\left(\varphi\right)
\end{eqnarray}
being $\hat{z}_{a i}$ and $z_{i a}$ normalization constants; here $\varphi_{j\in \partial a\setminus i}$ are the angles of the variable nodes linked to functional node $a$ other than $i$ and $\psi$ is the weight of the
functional node.  We will consider four-body interaction terms, such that
for an incoming message $a\to i$ the functional node $a$ has $3$  
connections $j_k$, $k=1,2,3$, other than $i$.
For the 4-$XY$ model $\psi$ reads:
\begin{eqnarray}
\psi( \{ \varphi_{i\in \partial a}\}|J_a)\equiv 
e^{\beta J_a 
	\cos \left(\varphi_{i_1} - \varphi_{i_2} + \varphi_{i_3} - \varphi_{i_4} \right)}
\label{eq:weight}
\end{eqnarray}
Eqs. (\ref{eq:uf},\ref{eq:update_variab}) are known as Replica Symmetric Cavity Method (RSCM) equations or Belief-Propagation (BP) equations.  Once the marginals are known, we can 
evaluate the free-energy and the order parameters, such as $q_{y y} = \langle \sigma_{y}\sigma_{y}\rangle$.
 The free energy reads\cite{Mezard09}:
\begin{eqnarray}
\label{eq:rs_free_en}
F\left(\underline{\nu},\underline{\hat{\nu}}\right) &= &\sum_{a=1}^M F_a + \sum_{i=1}^N F_i
-\sum_{(ia)\in E} F_{ia}
\\ \nonumber
F_a &\equiv& -\frac{1}{\beta} \log z_{cs} \left(\{\nu_{\cdot  a}\}, J \right)
\\ \nonumber 
F_i &\equiv &-\frac{1}{\beta} \log z_s \left( \{\hat{\nu}_{\cdot  i}\} \right)
\\ \nonumber
F_{ia} &\equiv& -\frac{1}{\beta} \log   z_l\left(\nu_{ia},\hat{\nu}_{a i}\right)
\end{eqnarray}
where $E$ is the set of all edges in the graph connecting variable and functional nodes;
 $z_{cs}$, $z_s$ and $z_l$ are: 
\begin{eqnarray}
z_{cs}\left(\{\nu_{ia}\}_{i \in \partial a}, J\right) & = & \int \prod_{k=1}^4 d \varphi_{i_k} \nu_{i_k a}(\varphi_{i_k})
\label{eq:z_c}\\
&& \qquad \qquad \times  \psi(\{ \varphi_{i_k\in \partial a}\}|J )
\nonumber
\\ 
z_s\left(\{\nu_{b i}\}_{b \in \partial i}\right) & = & \int_0^{2\pi} d \varphi \prod_{b=1}^c \hat{\nu}_{bi}(\varphi) \label{eq:z_s}\\
z_l\left(\nu_{ia},\hat{\nu}_{ai}\right) & = & \int d \varphi ~\nu_{ia}(\varphi)~ \hat{\nu}_{ai}(\varphi) \label{eq:z_l}
\end{eqnarray}
where $c$ in Eq. \eqref{eq:z_s} indicates the connectivity of node $i$; the subscripts $s$, $cs$, and $l$ stand for site, constraint
and link contribution respectively.\\ 
We are interested on ensembles of random factor graphs. In this case, Eqs. (\ref{eq:uf},\ref{eq:update_variab}) become equalities among distributions\cite{Mezard09,Mezard01} and the terms in Eq. \eqref{eq:rs_free_en}  are averaged over different realization of the graphs.\\
\indent 
The assumptions of the RSCM, i.e., the independence of messages coming from neighbours nodes, may fail if correlations do not decrease exponentially.
	With the one step replica symmetry breaking ansatz we can overcome this hypothesis assuming  a particular structure of the phase space: the states are organized in apart clusters, and, within each one of them, the assumptions of the RSCM are correct\cite{Mezard09}. Pure states occur with probability depending on its free-energy density, cf. Eq. 
	\eqref{eq:weight_sec}. By means of this measure over the pure states, we study the system introducing an auxiliary statistical-mechanical problem.
 The joint distribution induced by the weights $w_n(x)$ on the messages $\{\nu,\hat{\nu}\}$ takes the form:   
\begin{eqnarray}
 \mathcal{M}_x(\underline{\nu},\underline{\hat{\nu}})&&
 \propto e^{-x\beta  F\left(\underline{\nu},\underline{\hat{\nu}}\right)} 
 \label{eq:m_total}
 \\
&& \times \prod_{a\in M} \prod_{i \in \partial a} \mathbb{I}\left[
\hat{\nu}_{a i} = \hat{f}\left(\{\nu_{ja}\}_{j \in \partial a \setminus i},J_a\right)\right] 
\nonumber
\\
&&
\times \prod_{j \in N} \prod_{b \in \partial j} 
\mathbb{I}\left[\nu_{j b} = f\left(\{\hat{\nu}_{a j}\}_{a \in \partial j \setminus b}\right)\right]
\nonumber
\end{eqnarray}
The identity functions, $\mathbb{I}$, assure that, within each state, the messages, 
$\hat{\nu}_{a i}$ and $\nu_{jb}$, satisfy the RSCM  Eqs. (\ref{eq:uf},\ref{eq:update_variab}); indeed,  the functions
 $f$
and $\hat{f}$ 
are, respectively, 
 \begin{eqnarray}
\label{eq:f_and_hat_f}
  f\left(\{\hat{\nu}_{a j}\}_{a \in \partial j \setminus b}\right) &=& 
\frac{1}{z_{j b}} \prod_{a \in \partial j \setminus b} \hat{\nu}_{a j}
\\
\nonumber
 \hat{f}\left(\{\nu_{ja}\}_{j \in \partial a \setminus i},J_a\right)
 &=& \frac{1}{\hat{z}_{a i}} \int \prod_{j \in \partial a \setminus i} 
 d \varphi_j \nu_{j a}(\varphi_j) 
\\
&&  \hspace*{1cm}\times \psi(\{ \varphi_{j\in\partial a\setminus i}, \varphi_i\}|J_a )
\nonumber
 \end{eqnarray}
where $z_{j b}$ and $\hat{z}_{a i}$
are the normalization constants
  \begin{eqnarray}\label{eq:z}
   z_{j b}  &=&  \int d \varphi \prod_{a \in \partial j \setminus b} \hat{\nu}_{a j} (\varphi)
\\
 \label{eq:zhat}
\hat{z}_{a i} &=& \int d \varphi_i \prod_{j \in \partial a \setminus i} 
   d \varphi_{j} \nu_{j a}(\varphi_{j})\\ \nonumber
  &&  \hspace*{1.5cm}\times
   \psi(\{ \varphi_{j\in \partial a \setminus i},\varphi_i\}|J_a)
  \end{eqnarray}
The RSCM is introduced starting from the joint probability distribution of the variable nodes, $P(\varphi_1,\dots,\varphi_N)$. Indeed, through $P(\varphi_1,\dots,\varphi_N)$, the system is mapped on a graph 
and Eqs. (\ref{eq:uf},\ref{eq:update_variab}) are used as update rules to find the stationary marginal distributions of the variable nodes.
 When we move to random graphs, we are interested on the distributions 
of the messages, $P(\nu)$,$Q(\hat{\nu})$, that take into account the random environment around the nodes.  In this case, as well, from the joint probability distribution,  Eq. \eqref{eq:m_total},  
 we introduce a new graphical model whose variables are the set of messages $\{\nu,\hat{\nu}\}$. Then, we use the message passing algorithms to find their stationary marginal distributions. To be more clear, for each edge $\left(i~a\right)$ in a given realization, we will have two distributions,  $P(\nu_{i a})$ and $Q(\hat{\nu}_{a i})$, yielding the probability of having the messages $\nu_{i a}$ and $\hat{\nu}_{a i}$.   
As underlined above, we are interested on ensembles of random graphs. In this case then, we obtain a system of equations for the distributions $\mathcal{P}(\underline{\nu})$ and
$\mathcal{Q}(\underline{\hat{\nu}})$ of the distributions $P(\nu_{i a})$ and $Q(\hat{\nu}_{a i})$, respectively. We obtain\cite{Mezard09,Mezard01}:
 \begin{eqnarray}
 P(\nu) & \stackrel{d}{=} & \frac{1}{\mathcal{Z}}\mathbb{E}_{c}\int \prod_{b =1}^{c-1} d Q_b(\hat{\nu}_b) %
 \mathbb{I}\left[\nu = f\left(\{\hat{\nu}_b\}\right)\right] \left[z\left(\{\hat{\nu}_b\}\right)\right]^x 
 \nonumber
 \\ \label{eq:final_p}\\
 Q(\hat{\nu}) & \stackrel{d}{=} & \frac{1}{\hat{\mathcal{Z}}}\mathbb{E}_{J} \int \prod_{i =1}^{k-1} d P_i(\nu_i) %
 \mathbb{I}\left[\hat{\nu} = \hat{f}\left(\{ \nu_i\},J  \right)\right] \nonumber
\\ &&\qquad\qquad\qquad\qquad \qquad\times ~ \left[\hat{z}\left(\{ \nu_i\}, J\right)\right]^x
 \label{eq:final_q}
 \end{eqnarray}
 where $\mathbb{E}_c$ denotes the expected value over the distribution of the random 
 connectivity of variable nodes
 and $\mathbb{E}_J$ denotes the average over the values of the disordered interaction couplings. 
The normalization factors $z$ and $\hat{z}$, cf. Eqs. (\ref{eq:z}, \ref{eq:zhat}), to
the power $x$ enter as weight in
Eqs. (\ref{eq:final_p}, \ref{eq:final_q}), where we recall that $x$ is the RSB parameter, cf. Eq. \eqref{eq:weight_sec}; these terms come from the
weight  $w_n$. It is important to keep in mind the two level of randomness: one is due to the appearance of many pure states, the other one to the randomness of the environment around the nodes.  
 Like for  simple message passing, $\mathcal{Z}$ and $\hat{\mathcal{Z}}$ appearing in the message distribution equations are normalization constants.
The free-energy of what is called ``the replicated'' system is given by:
\begin{eqnarray}\label{eq:total_f}
-\beta \mathscr{F}(\beta,x) &=& 
\mathbb{E}_{\{Q(\hat{\nu})\},c} \log \mathcal{Z}_s\left(\{ Q(\hat{\nu}) \}\right) 
\\
\nonumber
&&+ %
\alpha \mathbb{E}_{\{P(\nu)\},J} \log \mathcal{Z}_c\left(\{P(\nu)\},J\right) 
\\
\nonumber
&&- n_l \mathbb{E}_{P(\nu),Q(\nu)}
 \log \mathcal{Z}_l\left(P(\nu),Q(\hat{\nu})\right)   
\end{eqnarray} 
where $\mathbb{E}$ indicates the expectation over the variables in
the indexes, $\alpha$ is the expected number of functional nodes per
variable node and $n_l$ is the expected number of edges per
variable.  The site, constraint and link
contributions are:
\begin{equation}\label{eq:site_contribution}
\mathcal{Z}_s = \int \prod_{b =1}^{c} d Q_b(\hat{\nu}_b) \left[z_s\left( \{\hat{\nu}_b \} \right)\right]^x
\end{equation} 
\begin{equation}\label{eq:c_contribution}
\mathcal{Z}_{cs} = \int \prod_{i =1}^{k} d P_i(\nu_i) \left[z_{cs}\left(\{ \nu_i\}, J\right)\right]^x
\end{equation}
\begin{equation}\label{eq:l_contribution}
\mathcal{Z}_l = \int d Q(\hat \nu) ~d P(\nu) \left[z_l\left(\nu,\hat{\nu}\right)\right]^x
\end{equation}
with $z_{cs}$, $z_s$ and $z_l$
 given by Eqs. (\ref{eq:z_c}, \ref{eq:z_s}, \ref{eq:z_l}), but now $J$ is a random variable.
The internal free-energy of the pure states, $\phi_{\rm int}$, is related to
the replicated free energy $\mathscr{F}(\beta,x)$ by a derivative with
respect to $x$: $\phi_{\rm int} = \partial{\mathscr{F}(\beta,x)}/\partial x$.
  We, thus,
obtain:
\begin{eqnarray}\label{eq:phi_int}
 -&\beta&\phi_{\rm int}(\beta,x)  = 
\\
\nonumber
&& \mathbb{E}_{\{Q(\hat{\nu})\}} \Biggl[\frac{1}{\mathcal{Z}_s}
\int \prod_{k =1}^{c} d Q_k(\hat{\nu}_k) \left[z_s\left( \{\hat{\nu}_k \} \right)\right]^x%
 \\
\nonumber
&&\qquad\qquad\qquad\qquad\qquad\qquad \times \log{z_s\left(\{\hat{\nu}_k \}\right)}
 \Biggr]  \\
\nonumber  &&  + \alpha \mathbb{E}_{\{P(\nu)\}} \Biggl[
 \frac{1}{\mathcal{Z}_{cs}}\int \prod_{i =1}^{4} d P_i(\nu_i) 
\left[z_{cs}\left(\{ \nu_i\}, J\right)\right]^x
\\
\nonumber
&&\qquad\qquad\qquad\qquad\qquad\qquad \times  \log{z_{cs}\left(\{ \nu_i\}, J\right)} \Biggr] 
\\
 && - n_l  \mathbb{E}_{P(\nu),Q(\nu)}%
 \Biggl[\frac{1}{\mathcal{Z}_l}\int d Q(\hat\nu)~ d P(\nu) \left[z_l\left(\nu,\hat{\nu}\right)\right]^x 
 \nonumber
 \\
&&\qquad\qquad\qquad\qquad\qquad\qquad \times \log{z_l\left(\nu,\hat{\nu}\right)}\Biggr] 
 \nonumber
\end{eqnarray}
The complexity, $\Sigma(\phi_{\rm int})$, related to the logarithm of the number of pure states with free-energy $\phi_{\rm int}$, can be evaluated through the Legendre transform:
\begin{equation}
\Sigma(\phi_{\rm int})=\beta\left(x \phi_{\rm int} - \mathscr{F}(\beta,x)\right) 
\end{equation}
The distributional Eqs.  (\ref{eq:final_p},\ref{eq:final_q}) depend on
the parameter $x$ and one eventually finds a self-consistent solution
for any value of $x \in [0,1]$. The total free energy
$\mathscr{F}(\beta,x)$, cf. Eq. (\ref{eq:total_f}), depends, as well, on $x$: in order to describe
the thermal equilibrium distribution we must determine the proper
value of $x$ by maximizing $\mathscr{F}(\beta,x)$\cite{Mezard09,Mezard87,Mezard01}. Let us analyse more in
details the two possible scenarios in which solutions cluster and,
consistently, replica symmetry is broken.

\subsubsection*{The dynamic 1RSB phase}
In this scenario the 1RSB solution displays $x=1$. From
a physical point of view, this is a consequence of the fact that the
internal free-energy
\textcolor{black}{is in  the dominion $[\phi_{\rm min},\phi_{\rm max}]$}    
where the complexity $\Sigma(\phi)$ is strictly positive.  We will
show that in this case the total free-energy $\mathscr{F}(\beta, 1)$
is equal to the paramagnetic one, {\em even if} it is obtained from a
superposition of an exponential number of pure states, each one with
free-energy $\phi_{\rm{int}}$ larger than the paramagnetic one.  This phase is also known as
dynamic $d-1RSB$ phase. We will show the behaviour of the previously introduced wave system for this case in Sec. \ref{sec:res}.

\subsubsection*{The static 1RSB phase}
In this scenario the system still decomposes into a convex combination of pure states
but a certain number of states acquire substantially more weight than all the
others: the probability measure condensates into this subset of
states growing sub-exponentially with the size (i.e., zero complexity)\cite{Mezard01}. The region $x=1$ becomes an unphysical solution of
negative complexity $\Sigma(\phi)$. 
The correct probability distributions
are obtained with $x=x^* \in [0,1)$.

We now concentrate on the first case,  where the maximum over $x$ is for
$x^*=1$. 

\subsection{1RSB phase with $x^*=1$}

In this section we determine the phase diagram of the $XY$ model with non-linearly interacting spins looking at the complexity function $\Sigma(\phi)$ when $x^*=1$. 
Another possible approach to determine critical points in linear systems, as developed in \cite{0305-4470-38-39-001,PhysRevE.72.066127} for $XY$ and Heisenberg spins\footnote{Heisenberg spins indicates unit vector in $3D$ systems, as well as $XY$ spins are unit vectors in $2D$ systems.}, is to look for a critical temperature through a bifurcation analysis: $P\left(\nu(\varphi)\right)$ is expanded around the paramagnetic solution, i.e.,  $P\left(\nu(\varphi)\right)=1/(2 \pi)+\Delta(\varphi)$, and the RSCM Eqs. (\ref{eq:uf},\ref{eq:update_variab}) are expanded as well around the solution 
$\Delta(\varphi)=0$; a linear update rule is then obtained for $\Delta(\varphi)$ and one can then analyse if $\Delta(\varphi)$ departs from zero. However, this procedure cannot be applied to the non-linear case, since one obtains a non-linear update rule and the behaviour of $\Delta(\varphi)$ cannot be determined analytically.\\
In the rest of this section we will introduce the variables and the order parameters that we analysed to study possible 1RSB solutions. The details of their derivation, are reported in Appendix \ref{app:rsb_m_1}.\\
Evaluating $\Sigma(\phi)$ at $x^*=1$ introduces a simplification in Eqs. (\ref{eq:final_p},\ref{eq:final_q}). 
To understand this result for a moment let us go back to one specific realization of a graph. Let us define the average messages, $\bar \nu_{ib}$,
$\bar{\hat{ \nu}}_{b i}$, over the possible states $n=1,\ldots, \mathcal{N}_{\rm states}(\phi)$ as:
\begin{equation}\label{eq:nu_av_first}
\bar \nu_{i b}  =  \int d P_{i b}(\nu) ~\nu \, ; \qquad %
\bar{\hat\nu}_{b i}  =  \int d Q_{b i}(\hat{\nu})~ \hat{\nu}
\end{equation}
If we look at a single factor graph taken from the ensemble of random factor
graphs we are considering, Eqs. (\ref{eq:final_p},\ref{eq:final_q})
for the specific distributions of the messages $\nu_{i b}$
and $\hat{\nu}_{b i}$ are:
\begin{eqnarray}\label{eq:specific_p}
P_{i  b}(\nu) & = & \frac{1}{\mathcal{Z}_{i  b}}\int \prod_{a \in \partial i \setminus b}%
 d Q_{a  i}(\hat{\nu}_a) \\ \nonumber  
&&\qquad \times 
 \mathbb{I}\left[\nu = f\left(\{\hat{\nu}_a\}\right)\right] \left[z\left(\{\hat{\nu}_a\}\right)\right] \\ \label{eq:specific_q} 
Q_{b i}(\hat{\nu}) & = &  \frac{1}{\hat{\mathcal{Z}}_{b i}}%
\int \prod_{j \in \partial b \setminus i} d P_{j b}(\nu_j) \\ \nonumber
&&\qquad \times \mathbb{I}\left[\hat{\nu} = \hat{f}\left(\{ \nu_j\},J  \right)\right] \left[\hat{z}\left(\{ \nu_j\}, J\right)\right] 
\end{eqnarray}
Through Eqs. (\ref{eq:z},\ref{eq:zhat}),
we can see that for this case $x^*=1$, the normalization constants,
$\mathcal{Z}_{i  b}$ and $\hat{\mathcal{Z}}_{b
  i}$, can be expressed in terms of the averages $\{\bar\nu,
\bar{\hat\nu}\}$:
\begin{equation}\label{eq:normalizations_const_z}
\begin{split}
\mathcal{Z}_{i  b} & = \int  \prod_{a \in \partial i \setminus b} d Q_{a  i}(\hat{\nu}_a) %
 z\left(\{\hat{\nu}_a\}\right)\\ 
 & = \int d \varphi_i  \prod_{a \in \partial i \setminus b} \int 
\left[d Q_{a i}(\hat{\nu}_a) ~\hat{\nu}_{a  i}(\varphi_i)\right] \\
& = \int d \varphi_i \prod_{a \in \partial i \setminus b} \bar{\hat\nu}_{a  i}(\varphi_i)  
\end{split}
\end{equation}
\begin{eqnarray}
\label{eq:normalizations_const_z_hat}
\hat{\mathcal{Z}}_{b  i} & =&\int \prod_{j \in \partial b \setminus i} d P_{j  b}(\nu_j)~ \hat{z}\left(\{ \nu_j\}, J_b\right)\\
\nonumber
& =&\int d \varphi_i  \left[\prod_{k=1}^3
d \varphi_{j_k} d P_{j_k  b}(\nu_{j_k}) ~\nu_{{j_k}  b}\right] 
\\
\nonumber
&&\qquad\qquad\qquad\qquad \times ~ e^{\beta J \cos \left(\varphi_{j_1} - \varphi_{j_2} + \varphi_{j_3} - \varphi_{i} \right)} \\
 \nonumber 
 &=& \int d \varphi_i \prod_{k=1}^3  d \varphi_{j_k} ~\bar\nu_{{j_k}  b} \quad
 e^{\beta J \cos\left(\varphi_{j_1} - \varphi_{j_2} + \varphi_{j_3} - \varphi_{i} \right)}          
\end{eqnarray}
The same can be observed for $\mathcal{Z}_s$, $\mathcal{Z}_{cs}$ and
$\mathcal{Z}_l$. 
  Moreover, from
Eqs. (\ref{eq:specific_p},\ref{eq:specific_q}), it can be seen that
$\bar \nu_{i  b}$ and $\bar{\hat\nu}_{i  b} $
satisfy the RSCM Eqs. (\ref{eq:uf},\ref{eq:update_variab}).\\
Moving to random graphs, 
 we can conclude
that, when the problem is correctly described by $x^*=1$, the
distributions of the average messages $\bar\nu$ and $\bar{\hat\nu}$,
which we will indicate with $\bar P$ and $\bar Q$, are
solutions of the RSCM equations and, in the thermodynamic limit, the
RS predictions are correct: the total free-energy $ \mathscr{F}(\beta,1)$ is 
equal to the  RS one. It, though, consists of {\em various contributions} due
to the fact that the thermodynamic phase is fragmented into many metastables
states,  each one with a free energy $\phi_{\rm{int}}$ larger than the RS one:
 \begin{equation}\label{eq:compl}
  \mathscr{F}(\beta,1) = \phi_{\rm int} - T \Sigma(\phi_{\rm int}) 
 \end{equation}
 Let us suppose that we know the RS results. Either because we
 know that a given solution, e.g., the paramagnetic solution, is the
 correct solution or because we have run the RS algorithm up to
 convergence. 
  We would like to simplify Eqs.(\ref{eq:final_p},\ref{eq:final_q})
 exploiting our knowledge that in this case the distributions of
 $\bar\nu$ and $\bar{\hat\nu}$ satisfy the RS predictions.  To this
 purpose let us define:
\begin{eqnarray}
R_{\varphi}(\nu) & = & \frac{\nu(\varphi)
  P(\nu)}{\bar\nu(\varphi)} \label{eq:r_nu} 
 \label{eq:R}
 \\
   \hat{R}_{\varphi}(\hat{\nu})
  & = & \frac{\hat{\nu}(\varphi)
  Q(\hat{\nu})}{\bar{\hat\nu}(\varphi)}
   \label{eq:r_hat_nu}
\end{eqnarray}
Starting from Eqs. (\ref{eq:final_p},\ref{eq:final_q}) we can obtain the distributional equations(see Appendix \ref{app:rsb_m_1}):
\begin{eqnarray} 
R_\varphi(\nu) & \stackrel{d}{=} & \Biggl\{ \int \prod_{b =1}^{c-1} d \hat{R}_{\varphi}(\hat{\nu}_{b})%
\mathbb{I}\bigl[\nu = f\left(\{\hat{\nu}_{b}\}\right)\bigr] \Biggr\} \label{eq:r_update} 
\\  
\hat{R}_{\varphi}(\hat{\nu}) & \stackrel{d}{=} & \mathbb{E}_{J} \Biggl\{%
\int \prod_{j =1}^{3}  d \varphi_j ~d R_{\varphi_j}(\nu_j)  \pi\bigl(\{\varphi_j\}|\varphi;J\bigr)\nonumber
\\
&&\qquad\qquad\times
\mathbb{I}\bigl[\hat{\nu} = \hat{f}\left(\{ \nu_{ j}\};J  \right)\bigr] \Biggr\} \label{eq:r_hat_update}
\end{eqnarray} 
where we have defined:
\begin{equation}
\pi\bigl( \{\varphi_{j_k}\}| \varphi; J_b \bigr) =
\frac{
	\prod_{j_k}{\bar\nu}_{j_k}(\varphi_{j_k})\times\psi(\{\varphi_{j_k}\},\varphi |J_b)}
{\int \prod_{j_k} d
	\varphi_{j_k} \bar\nu_{j_k}(\varphi_{j_k}) \times \psi(\{\varphi_{j}\},\varphi |J_b)}
\label{eq:pi_given_phi}
\end{equation}
that can be seen as the probability of a configuration $\{\varphi_{j_k}\}$ given $\varphi$ and  $J_b$.
In Eq. \eqref{eq:r_update},  for simplicity, we have considered the case of a random regular
factor graphs, in which the connectivity $c$ of each variable node is
constant.\\  
To numerically find the solutions of 
 Eqs. (\ref{eq:r_update}, \ref{eq:r_hat_update}) we can adopt an iterative scheme.  
Considering Eq. \eqref{eq:r_update}, $\hat{R}^{(t)}_{\varphi}(\hat{\nu})$ is updated through the following steps:
\begin{enumerate}
\item Let $J$ be drawn according to its distribution. Given $\varphi \in
  [0,2 \pi)$ and $J$, generate $\{\varphi_{j_1},\varphi_{j_2},\varphi_{j_{3}}\}$ according
      to $\pi\bigl(\{\varphi_j\}|\varphi;J\bigr)$.
\item Take the message marginals $\nu_1,\nu_2,\nu_{3}$ with
  distributions
  $R^{(t)}_{\varphi_{j_1}},R^{(t)}_{\varphi_{j_2}},R^{(t)}_{\varphi_{j_3}}$.
  Using the Population Dynamics Algorithm\cite{Mezard09}, this can be achieved
  considering each one of the $R_{\varphi_j}$ as a population of $N_{\rm pop}$
  messages. The $\nu_j$'s will then be taken uniformly at random from
  the $N_{\rm pop}$ messages.
\item Considering Eq. \eqref{eq:r_hat_update}, $\hat{\nu} =
  \hat{f}\left(\{ \nu_{ j}\};J \right)$ will be distributed as
  $\hat{R}^{(t+1)}_{\varphi}$. Numerically, as for the $R_{\varphi_j}$,
  $\hat{R}^{(t)}_{\varphi}$ is a population of $N_{\rm pop}$ distributions. Then, the new
  $\hat{\nu}$ will update one message among them, uniformly at random. Steps 1 to
  3 are repeated for all values of $\varphi$.
\end{enumerate}
To update Eq. \eqref{eq:r_update} we can proceed as follows.
\begin{enumerate}
\item Given $\varphi$, we take $c-1$ i.i.d messages,
  $\hat{\nu}_1,\dots,\hat{\nu}_{c-1}$ extracted with probabilities
  $\hat{R}^{(t)}_{\varphi}$;
\item Then, $\nu = f\left(\{\hat{\nu}_{b}\}\right)$ has distribution
  $R^{(t+1)}_\varphi$. As before, $\nu$ will update one of the $N_{\rm pop}$
  messages representing the distribution $R^{(t)}_\varphi$. Again,
  steps 1 and 2 have to be repeated for all values of $\varphi$.
\end{enumerate}
 Once  the stationary
distribution $\{\hat{R}^{(\infty)}_{\varphi},R^{(\infty)}_{\varphi}\}$ are computed we
can evaluate the order parameters. In spin-glass systems good order parameters are the inter and intra  state overlaps, $q_0$ and $q_1$. This latter is also known as the Edward-Anderson parameter. They read:
\begin{eqnarray}\label{eq:t_q_0}
q_0 & =& 
\left(\int d \varphi ~\bar\mu(\varphi)~ \cos{(\varphi)} \right)^2 + \left(\int d \varphi ~\bar\mu(\varphi) ~\sin{(\varphi)} \right)^2
\nonumber 
\\
\end{eqnarray}
depending only on the RS solutions, and
\begin{eqnarray}
q_1 & =&
\label{eq:t_q_1}
\int d \varphi ~\bar\mu(\varphi) \cos{(\varphi)} \\
\nonumber
&&\qquad\qquad\times
\mathbb{E}_{R_\varphi(\mu)}
\biggl[\int dR_{\varphi}(\mu)
\int d \varphi ~\mu(\varphi) \cos{(\varphi)} \biggr]
\\
&&+\int d \varphi ~\bar\mu(\varphi) \sin{(\varphi)} 
\nonumber
\\
\nonumber
&&\qquad\qquad\times
\mathbb{E}_{R_\varphi(\mu)}
\biggl[\int dR_{\varphi}(\mu)
\int d \varphi ~\mu(\varphi) \sin{(\varphi)} \biggr]  
\end{eqnarray}
Where, in analogy with $R_{\varphi}(\nu)$, cf. Eq. \eqref{eq:R}, we have defined: 
\begin{equation}
R_{\varphi}(\mu) = \frac{\mu(\varphi) P_\mu(\mu)}{\bar\mu(\varphi)}
\end{equation}
and $P_{\mu}$ is the distribution of the marginal probability
distributions $\mu(\varphi)$ of  variable nodes. $\mu_i(\varphi)$ of variable node $i$ is related to the messages coming from all the
neighbours of the node:
\begin{equation}\label{eq:mu}
\mu_i(\varphi) =\frac{\prod_{b \in \partial i} \hat{\nu}_{b  i }(\varphi)}{\int d \varphi \prod_{b \in \partial i} \hat{\nu}_{b  i }(\varphi)}
\end{equation}
As before, $\bar\mu(\varphi)$ indicates the average over the pure states. Eqs. (\ref{eq:t_q_0},\ref{eq:t_q_1}) are derived in App. \ref{app:rsb_m_1}. 
\subsection{$p$-clock model}
\begin{figure}[t!]
\includegraphics[width=.99 \columnwidth]{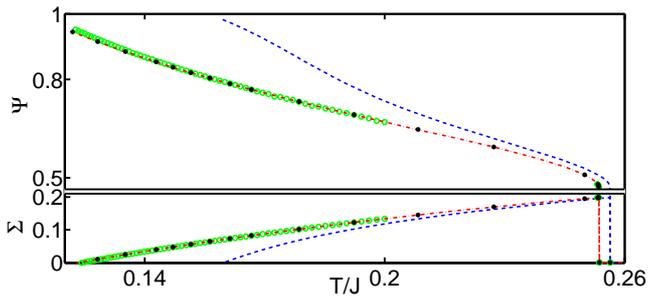}
\caption{Complexity, $\Sigma$, (bottom) and $\Psi\equiv \Psi_{\infty}$, (top) as functions of $T/J$ for different $p$-clock models on  random regular graphs with variable node connectivity fixed to $c=6$.  The dashed (blue) line and the empty (green) circles refer to $p=8$ and $p=12$, respectively. The dashed-dotted (red) line and the (black) stars refer to $p=16$ and $p=32$, respectively.
	At $T/J=T_d/J$ both $\Sigma$ and $\Psi$ show a discontinuous jump departing from zero  signaling a dynamical glass transition. As $T/J$ decreases below $T_d/J$, $\Sigma$ decreases towards zero. On the other hand, $\Psi$ increases. Below $T_K/J$, $\Sigma$ becomes negative and the correct solution is obtained with $x^*<1$. For the convergence of $T_d/J$ and $T_K/J$ as $p$ increases see Figs. \ref{fig:c5_Td} and \ref{fig:c5_TK}. We can see that down to $T=T_K/J$, no differences can be appreciated in double precision between $p=16$ and $p=32$.  The values obtained for $\Psi$ and $\Sigma$ are observed to be independent from $\rho$, i.e., the fraction of negative interactions. These results have been obtained with a population size of $N_{\rm pop} = 5 \cdot 10^5$.}
\label{fig:c6_Sigma}
\end{figure}

As we have seen for the ordered case,\cite{Marruzzo15} the method adopted  to look for numerical solutions of Eqs. (\ref{eq:r_update},\ref{eq:r_hat_update}) is to define  the $p$-clock model looking for a convergence to the $XY$-model for what concerns its critical 
behavior. We recall that in the $p$-clock model, the variable $\varphi$ can take $p$ values equally spaced in the $[0,2 \pi)$ interval:
\begin{equation}
 \varphi_m = \frac{2 \pi}{p} m \mbox{ with } m = 0, \dots, p-1
 \label{eq:pclock}
\end{equation}
The population dynamics algorithm for the $p$-clock model is reported in detail in Appendix \ref{app:pclock}.\\ 
The last point concern  the starting distributions, the $R^{(0)}_m$s. It was seen in Ref. \onlinecite{Mezard06} that the temperature of the dynamical glass transition coincides with the noise threshold of an associated reconstruction problem: the statistical physics problem admits a glassy phase if the related reconstruction problem is successful.  
Therefore, the 1RSB equations
have solutions different from the RS ones if and only if Eqs. (\ref{eq:r_update},\ref{eq:r_hat_update}), with initial conditions 
 $R^{(0)}_m(\nu) = \mathbb{I}(\nu_{m'} = \frac{p}{2 \pi} \delta_{m' m})$, do not converge to the trivial distributions $R^{(\infty)}_m(\nu) = \mathbb{I}(\nu = \nu_{RS})$, $\forall m \in \{0,\dots,p-1\}$, as
$t \rightarrow \infty$; where, for the $XY$ model,  $\nu_{RS}=1/(2 \pi)$.
In order to better understand the analysis that we will show in the following, we briefly sketch this result introducing the so-called \emph{tree-reconstruction problem} \cite{Mezard06}, which will also clarify the physical meaning of $R_{\varphi}(\nu)$ introduced in Eq. \eqref{eq:R}.\\
Consider a source generating a signal that propagates through the links of a tree-like network. 
The tree reconstruction deals with the problem of 
forecasting the signal at the source knowing the broadcasted signal at the end leaves 
 of the tree.
 Let us call $\underline{M}_{l}$ the configuration at distance $l$ from the root (in \cite{Mezard06} this is called the $l$th generation) and
let us define $\eta_l(\tilde{m})$ as the probability that the root has value $\tilde{m}$, i.e., it is pointing in the $\tilde{m}$ angle direction,  
given $\underline{M}_{l}$; this reads
\begin{equation}\label{eq:eta_def}
 \eta_l(\tilde{m}) = \mathbb{P}\{M_0=\tilde{m} | \underline{M}_l=\underline{m}_l\}
\end{equation}
Since $\underline{m}_{l}$ is chosen randomly according to the broadcast from the root to the $l^{\rm th}$ generation, $\eta_l(\tilde{m})$ is in general a random variable. 
Let us indicate with $R^{(l)}_{\tilde{m}}(\eta)$ the probability distribution of $\eta_l(\tilde{m})$, conditional to the broadcast problem being started with $M_0 = \tilde{m}$:
\begin{equation}
 R^{(l)}_{\tilde{m}}(\eta) = \mathbb{P}\{\eta_l(\cdot)=\eta(\cdot)|M_0=\tilde{m}\}
\end{equation}
Through $R^{(l)}_{\tilde{m}}(\eta)$ we can determine the probability that the reconstruction problem is successful:
$$
P_{\rm success} = \frac{1}{p} \sum_{\tilde{m}=0}^{p-1} \int d R^{(l)}_{\tilde{m}}(\eta)~ \eta_l(\tilde{m})
$$
where we have considered variables $\tilde{m}$ discrete taking values in $\{0,\dots,p-1\}$.\\
Parallel to $R^{(l)}_{\tilde{m}}(\eta)$, giving the distribution of $\eta_l(\cdot)$ conditional on the transmitted signal being equal to $\tilde{m}$, it is also interesting to consider the unconditional distribution of $\eta_l(\cdot)$. 
 Using the Bayes theorem we have:
\begin{equation}\label{eq:b}
 \mathbb{P}\{\eta_l=\eta|M_0=\tilde{m}\} = \frac{\mathbb{P}\{M_0=\tilde{m} | \eta_l=\eta\} \mathbb{P}\{\eta_l=\eta\}}{\mathbb{P}\{M_0=\tilde{m}\}}
\end{equation}
Using the definition of $\eta_l(\cdot)$, Eq. \eqref{eq:eta_def}, we can see that Eq. \eqref{eq:b} coincides with Eq. \eqref{eq:R}.\\ 
The behaviour of $R^{(l)}_{\tilde{m}}(\eta)$ as $l$ increases will give information about the possibility of actually reconstruct the broadcast signal $M_0$. Indeed, we can define  the reconstruction probability as the probability that the reconstruction is successful minus the probability of guessing uniformly: 
\begin{equation}\label{eq:psi}
 \Psi_l = \frac{1}{p} \sum_{\tilde{m}=0}^{p-1} \int d R^{(l)}_{\tilde{m}}(\eta) \left[ \eta_l(\tilde{m}) - \frac{1}{p}\right]
\end{equation}
 In terms of the above observable we can say that the problem is solvable if, in
the limit $l \rightarrow \infty$,
$\Psi_{\infty} > 0$.
The further step is to determine how we can compute the distribution $R^{(l)}_{\tilde{m}}(\eta)$ of the distributions of the root $\eta_l(m)$ given a boundary 
$\underline{M}_l=\underline{m}_l$. Using the tree-like structure of the problem, this can be done iteratively by a dynamical programming procedure that induces a recursion 
equation for  $R^{(l)}_{\tilde{m}}(\eta)$. It can be seen that this coincides with Eq. \eqref{eq:r_update}\cite{Mezard06}. 
The initial condition is then:
\begin{equation}\label{eq:init}
R^{(0)}_{\tilde{m}}(\eta) = \delta\left[\eta - \delta_{\tilde{m}}\right]
\end{equation}
As it is done in the RSCM, the idea is to use the recursion rule on diluted graphs that are locally tree-like. The reconstruction problem is then  mapped on the statistical mechanical problem and this latter  admits a glassy phase if, and only if, the corresponding reconstruction problem is solvable, i.e., $\Psi_{\infty} >0$. 
We take as initial condition Eq. \eqref{eq:init} and evaluate $\Psi_{\infty}$
 to identify the occurrence of a possible glassy phase.
 As a comparison, we have also evaluated the complexity  function from Eq. \eqref{eq:compl},
which is expected to jump from zero to a value $\Sigma > 0$ at the dynamical glass transition, $T_d$, where the phase space decomposes into an exponential number of pure states.
As expected, the two quantities $\Sigma$ and $\Psi$
 display a discontinuity right at the same reduced temperature $T/J \equiv T_d/J$, as shown in Fig. \ref{fig:c6_Sigma} for 
 various $p$-clock models on a bipartite Bethe lattice with $c=6$ and  $k=4$. The tree reconstruction probability $\Psi$ is, anyway, easier to control because it fluctuates less than $\Sigma$.
We notice that for the continuous variable case, $\varphi\in[0,2\pi)$, Eq. \eqref{eq:psi} is replaced by
\begin{equation}
\nonumber
\Psi_{t} = \frac{1}{2 \pi} \int d \varphi \int d R^{(t)}_\varphi(\eta) \left[\eta(\varphi) - \frac{1}{2 \pi} \right]
\end{equation}
\section{Phase Transitions in the $p$-clock models }
\label{sec:res}
\begin{figure}[t!]
\includegraphics[width=.99 \columnwidth]{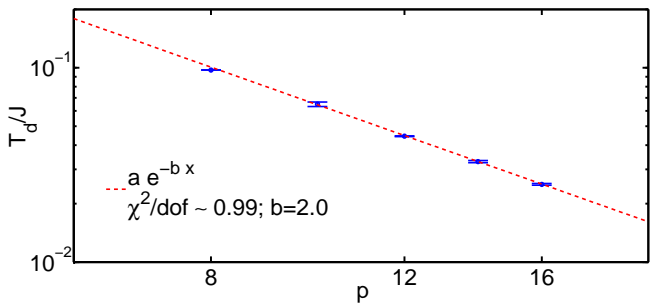}
\caption{Dynamical transition values, $T_d/J$, as a function of $p$ for  random graphs with node connectivity fixed to $c=4$. The data are show with their best fit. We can see that in the $p\rightarrow \infty$ limit,  i.e., the $XY$ model, no dynamical glass transition is expected for this value of connectivity and results indicating values of $T_d/J>0$ are  artifacts of $p<\infty$.}
\label{fig:c4_Td}
\end{figure}

\begin{figure}[t!]
\includegraphics[width=.99 \columnwidth]{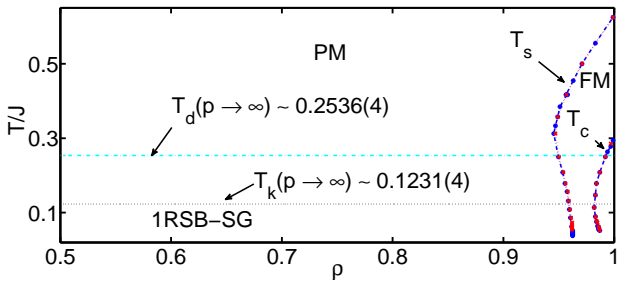}
\caption{Phase diagram obtained for random graphs with node connectivity fixed to $c=6$ in the $\rho$, $T/J$ plane. Three phases are found: the Paramagnetic phase (PM) corresponding to  the continuous wave phase  in the wave system, the Ferromagnetic phase (FM), i.e., the mode-locked  regime, and 1RSB Spin-Glass (SG) phase, i.e., the glassy  regime. The dashed-dotted (cyan) and dotted (black) lines show the dynamic, $T_d/J$, and static, $T_K/J$, transition lines, 
	respectively, as the number of negative interaction increases. $T_d/J$, as well as $T_K/J$,  does not depend on $\rho$. The (blue) circles and the (blue) dashed-dotted line on the right indicate the boundary between the FM and SG/PM phases obtained for $p=32$, while the (red) crosses and the (red) dotted line refer to the results for $p=16$. With $T_s$ we indicate the \emph{spinodal} line, i.e., the value of $T/J$ at which the FM solution appears; $T_c$ is the critical line and it indicates the value at which the FM solution becomes firstly stable. For $\rho \lesssim 0.95$ the FM solution vanishes.}
\label{fig:c6_PhDi}
\end{figure}

Eventually, we present the behaviour of the dynamic transition line and the static, Kauzmann-like, transition line  as the number $p$ of values  that the phase variables can take increases. We have considered random  graphs with different connectivities $c$. The results presented have been obtained with a population size of $N_{\rm pop} = 5 \cdot 10^5$ elements; for this value no size effects could be observed.\\
The borderline case to display any transition at all, at least for $p<\infty$, is when the connectivity of the variable nodes is equal to the connectivity of the function nodes, that is $4$ in the model introduced in Eq. \eqref{eq:H2}. In Fig. \ref{fig:c4_Td}
we show how a finite $T_d>0$ is an artifact of taking $\varphi$ discrete in the $p$-clock models. Indeed, increasing $p$,
$T_d$ tends to zero as a power law. Consequently, any $T_K<T_d$ also tends to zero for large $p$.
\begin{figure}[t!]
\includegraphics[width=.99 \columnwidth]{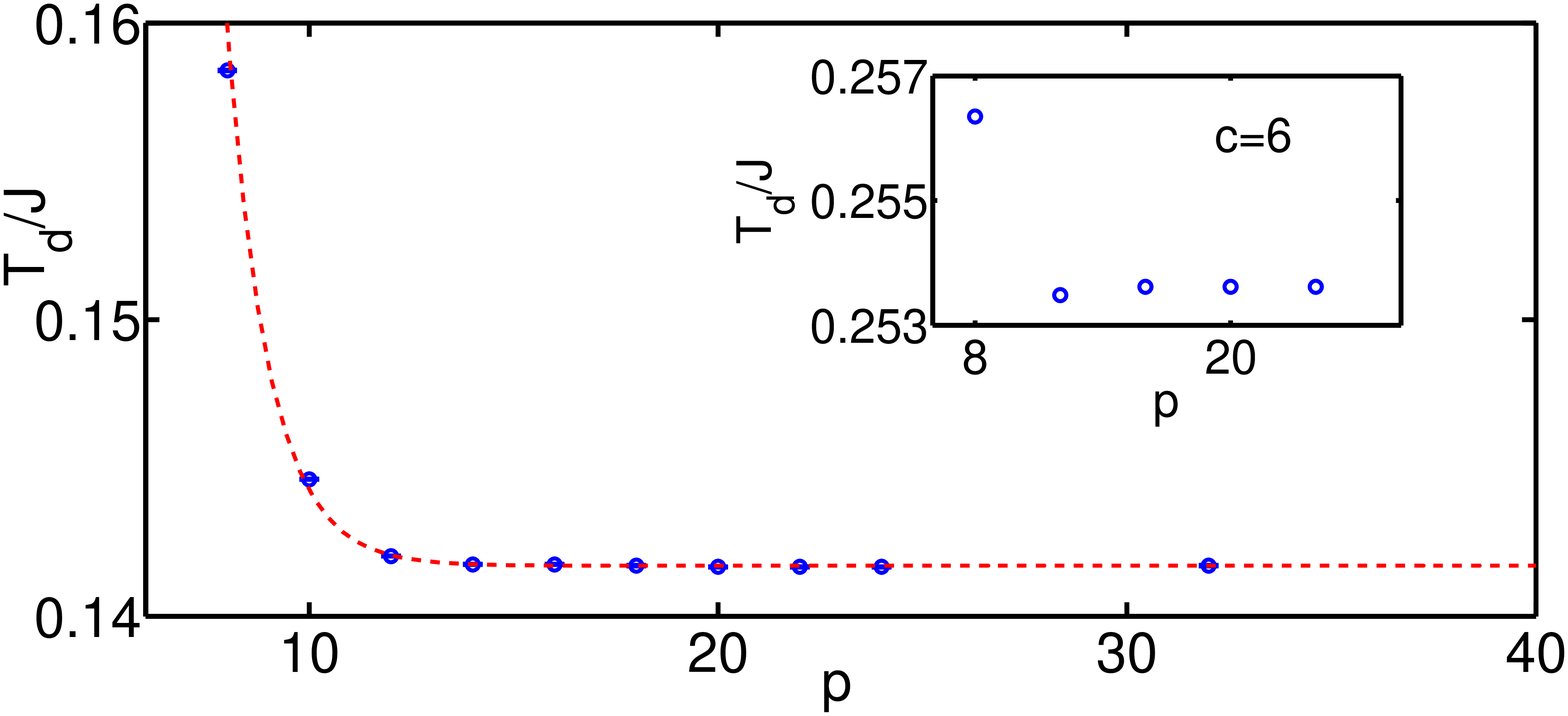}
\caption{Dynamical transition values, $T_d/J$ (open blue circles), as a function of $p$ with their best fit (dashed red line) obtained in random graphs with node connectivity fixed to $c=5$. The fitting functions is $a e^{-b  p}+T^{\rm{lim}}_d/J$ with $b=1.0$ and limiting value  $T^{\rm{lim}}_d/J= 0.141(7)$; $\chi^2/\rm{dof} = 0.99$. The inset shows  the results for random graphs with node connectivity fixed to $c=6$; the limiting value in this case is $T^{\rm{lim}}_d/J= 0.2536(4)$. We note that convergence is reached for a lower value of $p$ in respect to $c=5$.}
\label{fig:c5_Td}
\end{figure}
\begin{figure}[t!]
\includegraphics[width=.99 \columnwidth]{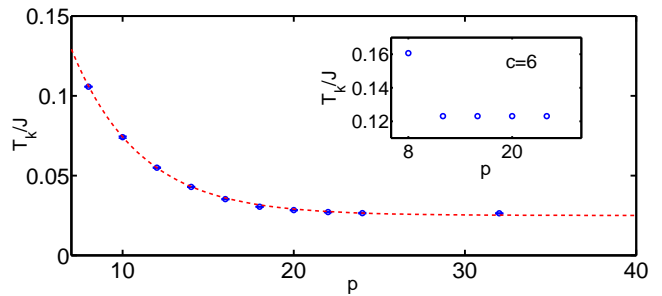}
\caption{Open blue circles refer to the values for the static (Kauzmann) transition temperature, $T_K/J$, plotted as  function of $p$ obtained in graphs with connectivity $c=5$. The inset  shows the results related to graphs with connectivity $c=6$. The results for $c=5$ are shown with their best fit,  dashed (red) line. The fitting function is: $a e^{- b p}+T^{\rm{lim}}_K/J$; results are: $b \sim 0.25$ and 
	$T^{\rm{lim}}_K/J = 0.026(3)$, $\chi^2/\rm{dof} = 0.99$. Concerning the inset for $c=6$, convergence to $T^{\rm{lim}}_K/J = 0.1231(4)$ is faster.}
\label{fig:c5_TK}
\end{figure}
Considering  larger  node connectivity values we observe a rapid convergence to asymptotic values of $T_d$ and $T_K$. This behavior is exemplified in Figs. \ref{fig:c5_Td} and  \ref{fig:c5_TK} for connectivity $c=5$, while the insets of these  figures show the results for $c=6$:
 as $c$ increases the convergence to the continuous $\varphi$ limit is faster. Indeed, at the latter case this is reached already for $p=12$ both for dynamic and static transitions. 
In the case $c=6$, we show the asymptotic, i.e., $p\rightarrow\infty$, phase diagram in Fig. \ref{fig:c6_PhDi}. Three phases are shown, each one of them both as stable and metastable: the incoherent wave (aka paramagnetic) regime, the totally phase coherent (aka ferromagnetic) regime and the disordered frozen regime in which all phases take a given fixed value but in random directions. The phase wave regime, with locked phases but overall zero magnetization, reported in Ref. \onlinecite{Marruzzo15} is also a solution but it always occurs as metastable and it is not reproduced here.

\begin{figure}[t!]
\includegraphics[width=.99 \columnwidth]{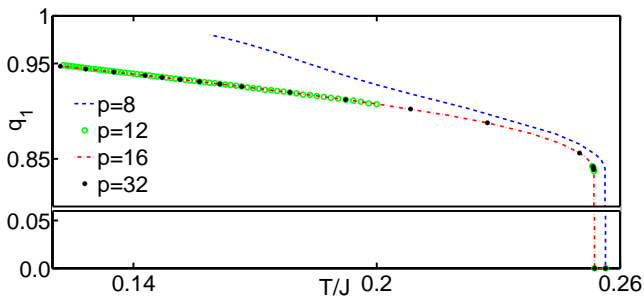}
\caption{Inter-state overlap, $q_1$, obtained for $c=6$. $q_1$ departs from zero at $T_d/J$ and increases. Different lines refer to different $p$-clock models (see the legend). As we can see  $p=16$ is already a good approximation 
	for the $p \rightarrow \infty$ limit. As $\Psi$ and $\Sigma$, values of $q_1$ are shown to be independent on $\rho$.}
\label{fig:c6_q1}
\end{figure}

At the dynamic transition, determined very precisely by the tree reconstruction probability, the self-overlap $q_1$ jumps to a non-zero value. The discontinuity is clearly seen for all $p$-clock models investigated, though  both its position, i.e., $T_d$,  and its magnitude in the glassy phase vary with $p$ for relatively small $p$ values, as shown in Fig. \ref{fig:c6_q1} for different $p$-clock models on random graphs with variable node connectivity $c=6$.

\section{Conclusions}
\label{sec:end}
In this work, we have investigated the $4$-body $XY$ model with quenched disordered interactions on bipartite random graphs. The system is highly diluted with the node participating in $\mathcal{O}(1)$ quadruplets and, by means of the Cavity Method, we have analyzed the possible phases from the ordered to the spin-glass case. To numerically implement the Cavity Method up to the 1-step Replica Symmetry Breaking, which is characterized by the presence of many pure states in the phase space and it is known to be the stable solution for nonlinear models, we have introduced the $p$-clock model representing, from $2$ to $\infty$,  a hierarchy of discretization for the $XY$ spins. Then, we have studied the dynamical and spinodal thermodynamic transition lines analyzing the free-energy, the complexity and the tree reconstruction functions
as both the distribution of the interaction couplings and the connectivity $c$ of the variable nodes are changed. The analysis of the convergence of the $p$-clock model to the
$XY$ model is performed down to temperature low enough to determine the asymptotic value of all relevant transition points
for different connectivity. The limiting values, as $p$ increases, of the complexity, the tree-reconstruction and the Edward-Anderson overlap $q_1$  are  analysed,
as well, in function of temperature. Our results indicate that, given the connectivity $c$, the dynamic and thermodynamic transitions are completely independent of the parameter $\rho$, Eq. \eqref{eq:rho}, giving the fraction of positive couplings. This property is clear from the horizontal lines in the $T-\rho$ plane in Fig. \ref{fig:c6_PhDi}. Moreover, looking at $\Sigma(\phi_{\rm int})$, $\Psi$ and $q_1$,
we can conclude that every unmagnetized solution of the self-consistent cavity equations is independent of $\rho$ and the spin-glass solution is also present when the ferromagnetic solution arises\footnote{We remember that this result was obtained also in \cite{0295-5075-55-4-465,1751-8121-44-18-185002} for the Ising model with nonlinearly interacting spins}; although, in this  case, the ferromagnetic solution is more favourable. Studying the limit $p\rightarrow \infty$,  our results indicate that the minimal connectivity to have a spin-glass solution for the $4$-$XY$ spin model at temperature above zero is $c=5$.\\
 The $XY$ model with pairwise, i.e., linearly, interacting
spins is well known in statistical mechanics,
 displaying important physical insights in different fields of condensed matter physics. In the present case, our interest on the many-body, i.e., non-linear, $XY$ model resides in optics, since it turns out to describe systems of interacting waves
with homogeneous distribution of the intensities. Disorder-induced frustration in the mode-couplings could effectively describe the effects of the scattering feedback in random lasers, where a spin-glass phase could be the source of recently observed anomalous spectral fluctuations in different identical experiments on the same sample \cite{Ghofraniha15, Gomes15, Pincheira15}.
\appendix
\section{1RSB Cavity Method with $x^*=1$: a simpler recursion}
\label{app:rsb_m_1}
In this appendix we will show how we can obtain the simpler recursion rules, Eqs. (\ref{eq:r_update},\ref{eq:r_hat_update}), starting from Eqs. (\ref{eq:final_p},\ref{eq:final_q}) 
knowing that the system is correctly described by states with $\phi_{\rm int} \in [\phi_{\rm min},\phi_{\rm max}]$, i.e., $x^*=1$.\\
Let us consider the single realization of a graph. Taking a link $i ~ b$, the messages $\nu^n_{i b}(\varphi)$ and $\hat{\nu}^n_{b i}(\varphi)$ will in general depend on the state $n$. We define the average messages,  $\bar \nu_{ib}$,
$\bar{\hat{ \nu}}_{b i}$, with Eqs. \eqref{eq:nu_av_first}. From Eq. \eqref{eq:specific_p}, using the identity function $\mathbb{I}\left[\nu = f\left(\{\hat{\nu}_a\}\right)\right]$, 
 we obtain:
\begin{equation}\label{eq:bar_nu}
\bar{\nu}_{ib}\left(\varphi\right)  =  \frac{1}{z_{i b}\left(\{\bar{\hat{\nu}}\}\right)} \prod_{a \in \partial i \setminus b} \bar{\hat{\nu}}_{a i} \left(\varphi\right) 
	\end{equation}
	here $z_{i b}\left(\{\bar{\hat{\nu}}\}\right)$ is again the normalization constant (Eq. \eqref{eq:z}) depending only on the $\bar{\hat{\nu}}_{a i}$ with $a \in \partial i \setminus b$.
	From Eq. \eqref{eq:specific_q}, using again $\mathbb{I}\left[\hat{\nu} = \hat{f}\left(\{ \nu_j\},J  \right)\right]$, we obtain: 
	\begin{eqnarray}\nonumber
	\bar{\hat{\nu}}_{b i}\left(\varphi\right) &=&  
	\frac{1}{\hat{z}_{b i}\left(\{\bar{\nu}\}\right)} \int_{0}^{2 \pi} \prod_{j \in \partial b \setminus i} 
	\left[d \varphi_{j} \bar{\nu}_{j b} \left(\varphi_{j} \right)\right] 
	\\  \label{eq:bar_hat_nu}
	&&\qquad\qquad\times 
	\psi(\{ \varphi_{j\in\partial b\setminus i}, \varphi\}|J_b)
	\end{eqnarray}
	as before, $\hat{z}_{b i}$ is the normalization constant (Eq. \eqref{eq:zhat}) depending only on $\bar{\nu}_{j b}$ with $j \in \partial b \setminus i$.  
	Eqs. (\ref{eq:bar_nu},\ref{eq:bar_hat_nu}) are the Belief Propagation equations, cf. Eqs. (\ref{eq:uf},\ref{eq:update_variab}). Considering ensemble of random graphs, these become equations among distributions: $\bar{P}(\nu)$ and $\bar{Q}(\hat{\nu})$ are self-consistent solutions of the RSCM equations. To derive Eqs. (\ref{eq:r_update},\ref{eq:r_hat_update}) we consider again an instance of a random graph. Given a link, $i ~ b$ we can then define:
	$$
	R^{i b}_{\varphi}(\nu) = \frac{\nu(\varphi) P_{i b}(\nu)}{\bar{\nu}_{i b}(\varphi)}
	$$ 
	similarly, we define $\hat{R}^{b i}_{\varphi}(\hat{\nu})$. We recall that $P_{i b}(\nu)$ describes the distribution of the message $\nu_{i b}(\varphi)$ among the different pure states $n=1,\dots,\mathcal{N}_{\rm{states}}(\phi)$. Through Eq. \eqref{eq:specific_p}, we can see that:
	\begin{eqnarray}
	\nonumber
		R^{i b}_{\varphi}(\nu) & = & \frac{1}{z_{i b}\left(\{\bar{\hat{\nu}}\}\right)\bar{\nu}_{i b}(\varphi)} \int \Biggl\{ \prod_{a\in \partial i \setminus b} \left[ d Q_{a i}(\hat{\nu}_{a}) \hat{\nu}_a(\varphi)\right]\Biggr.\\ \nonumber
		&&\Biggl. \mathbb{I}\left[\nu = f\left(\{\hat{\nu}_a\}\right)\right]\Biggr\}\\  \label{eq:r_inst}
		&=& \int \prod_{a\in \partial i \setminus b}  d \hat{R}^{a i}_{\varphi}(\hat{\nu}_{a}) \mathbb{I}\left[\nu = f\left(\{\hat{\nu}_a\}\right)\right]
	\end{eqnarray}  
	where we have  written $\nu(\varphi)$ using $\mathbb{I}\left[\nu = f\left(\{\hat{\nu}_a\}\right)\right]$ and we have simplified 
	$z_{ib}\left(\{\bar{\hat{\nu}}\}\right)\bar{\nu}_{i b}(\varphi)$ using 
	Eq. \eqref{eq:bar_nu}. Through the same steps, we can write:
	\begin{eqnarray}
	\nonumber
		\hat{R}^{b i}_{\varphi}(\hat{\nu}) & = & \Biggl\{\int \prod_{k=1}^3 \left(d \varphi_{k}\right) \prod_{j\in \partial b \setminus i}\biggl[\int d R^{j b}_{\varphi_{k}}(\nu_k) \bar{\nu}_{j b}(\varphi_k)\biggr]\Biggr.\\ \label{eq:r_hat_inst}
		& &\Biggl. \times \psi(\{\varphi_k\};\varphi,J_b)\mathbb{I}\left[\hat{\nu} = f\left(\{\nu_k\}\right)\right]\Biggr\}\\ \nonumber
		& & \times \Biggl\{\int \prod_{k=1}^3 \left(d \varphi_{k}\right) \psi(\left\{\varphi_k\right\};\varphi,J_b)\prod_{j\in \partial b \setminus i} \bar{\nu}_{j b}(\varphi_k)\Biggr\}^{-1}
	\end{eqnarray} 
	We can define:
	$$
	\pi(\{\varphi_k\}|\varphi;J_b) = \frac{\prod_{j\in \partial b \setminus i} \bar{\nu}_{j b}(\varphi_k)\psi(\{\varphi_k\};\varphi,J_b)}{z_{\pi}\left(\varphi,J_b\right)}
	$$
	where $z_{\pi}\left(\varphi,J_b\right)$ is the normalization:
	$$
	z_{\pi}\left(\varphi,J_b\right)=\int \prod_{k=1}^3 \left(d \varphi_{k}\right)\prod_{j\in \partial b \setminus i} \bar{\nu}_{j b}(\varphi_k)\psi(\{\varphi_k\};\varphi,J_b)
	$$ 
	Hence, $\pi(\{\varphi_k\}|\varphi;J_b)$ can be seen as the probability distribution of $\{\varphi_k\}$ given $\varphi$ and $J_b$. Considering ensemble of random factor graphs, Eqs. 
	(\ref{eq:r_inst},\ref{eq:r_hat_inst}) become equalities among distributions  (Eqs. (\ref{eq:r_update},\ref{eq:r_hat_update})).\\
	It can be seen that $\mathscr{F}(\beta,1)$, Eq. \eqref{eq:total_f}, depends only on the average messages, $\{\bar{\nu},\bar{\hat{\nu}}\}$, coinciding then with the RS free-energy, Eq. \eqref{eq:rs_free_en}. Indeed, when $x^*=1$,  $\mathcal{Z}_s$ of Eq. \eqref{eq:site_contribution} reads:
	\begin{eqnarray}
	\nonumber
	\mathcal{Z}_s &=&\prod_{b=1}^c \bigl(\int d Q_b(\hat{\nu}_b)\bigr) \int d \varphi \prod_{b=1}^c \hat{\nu}_b(\varphi)\\
	&=& \int d \varphi\prod_{b=1}^c  \bigl(\int d Q_b(\hat{\nu}_b) \hat{\nu}_b\bigr) = \int d \varphi \prod_{b=1}^c\bar{\hat{\nu}}_b
	\end{eqnarray}
	the same can be obtained for $\mathcal{Z}_c$ and $\mathcal{Z}_l$.\\
	The more common order parameters used to describe the 1RSB solutions are the intra-state overlap, $q_1$, and the inter-state overlap, $q_0$. The first describes the similarity among configurations belonging to the same pure state, the latter is the overlap among configurations belonging to different states. We expect $q_0<q_1$ if the replica symmetry breaking occurs, otherwise $q_0=q_1$. From their definitions, $q_0$ and $q_1$ can be evaluated through:
	\begin{eqnarray}\label{eq:q_0_step}
	q_0 & = & \mathbb{E}_{P_{\mu}} \Biggl\{ %
	\left(\int d P_{\mu}(\mu) \int d \varphi \cos{(\varphi)} \mu(\varphi) \right)^2 
	\\
	\nonumber
	&&\qquad\qquad +
	\left(\int d P_{\mu}(\mu) \int d \varphi \sin{(\varphi)} \mu(\varphi) \right)^2 
	\Biggr\} \\
	q_1 & = & \mathbb{E}_{P_{\mu}} \Biggl\{ \int d P_{\mu}(\mu) \Biggl[\left(\int d \varphi \cos{(\varphi)} \mu(\varphi) \right)^2 
	\\
	\nonumber
	&&\qquad\qquad + \left(\int d \varphi \sin{(\varphi)} \mu(\varphi) \right)^2\Biggr] \Biggr\} \label{eq:q_1_step}  
	\end{eqnarray}
	where $P_{\mu}$ is the distribution of the marginal probability
	distributions $\mu(\varphi)$, Eq. \eqref{eq:mu}. $P_{\mu}$ is a self-consistent solution of a 1RSB cavity equation similar to Eq. \eqref{eq:final_p} with $c-1$
	replaced by $c$.
	When $x^*=1$, as we did in Eqs. (\ref{eq:r_nu},\ref{eq:r_hat_nu}),
	we can define:
	\begin{equation}
	R_{\varphi}(\mu) = \frac{\mu(\varphi) P_\mu(\mu)}{\bar\mu(\varphi)}
	\end{equation}
	Substituting in Eq. (\ref{eq:q_0_step}) we obtain Eq. \eqref{eq:t_q_0}. 
 For $q_1$ we have:
 \begin{eqnarray}
 \nonumber
 q_1&=& \int d \varphi_1 ~d \varphi_2 \cos{(\varphi_1)} \cos{(\varphi_2)}\\
 \nonumber & & \times ~
 \mathbb{E}_{P_\mu} \int d P_\mu \frac{\mu(\varphi_1) \bar\mu(\varphi_1)}{\bar\mu(\varphi_1)} 
 \mu(\varphi_2)  + \cos \leftrightarrow \sin\\ \nonumber
 & =& \int d \varphi_1 \cos{(\varphi_1)}\bar\mu(\varphi_1)\\ \nonumber
 & & \times ~ \mathbb{E}_{R_{\varphi_1}} \int d R_{\varphi_1}(\mu)\bigl[\int d \varphi_2 \cos{(\varphi_2)} \mu(\varphi_2)\bigr]\\
 & & \quad +  \cos \leftrightarrow \sin
\end{eqnarray}
Through similar calculations, when $x^*=1$, we can also use $R_{\varphi}$ and $\hat{R}_{\varphi}$ to simplify the expression for $\phi_{\rm int}$, Eq.  \eqref{eq:phi_int}.
Indeed, starting from Eq. \eqref{eq:phi_int}, we obtain:
\begin{eqnarray}
\label{eq:phi_int_x_1}
&&-\beta \phi_{\rm int}(\beta) =
\\
\nonumber
&&
\mathbb{E}_{\{\hat{R}\},\{\bar{\hat\nu}\}} \Biggl[
\frac{\int d \varphi \prod_{b=1}^c \left(\bar{\hat\nu}_b 
	\int d \hat{R}_\varphi(\hat{\nu}_b) \right)%
	\log{z_s\left(\{\hat{\nu}_b\}\right)}}{z_s\left(\{\bar{\hat{\nu}}_b\}\right)} 
\Biggr]
\\
&&\quad
\nonumber 
+ 
\alpha \mathbb{E}_{\{R\},\{\bar\nu\}}
\Biggl[\frac{1}{z_{cs} \left(\{\bar\nu\},J\right)}
\\
\nonumber
&&\qquad\times
\int \prod_{j=1}^4  d \varphi_{j} \bar\nu_{j}(\varphi_{j}) 
d R_{\varphi_j} \psi(\{\varphi_j\} |J)
\log{z_{cs}\left(\{\nu_j\},J\right)} \Biggr] \\
\nonumber 
&& \quad - n_l \mathbb{E}_{\hat{R},R,\bar{\hat\nu},\bar\nu}\frac{\int d \varphi ~\bar\nu(\varphi) \bar{\hat\nu}(\varphi) \int d R_{\varphi} d \hat{R}_{\varphi}%
	\log{z_l\left(\nu,\hat{\nu}\right)}}{z_l\left(\bar\nu,\bar{\hat\nu}\right)}
\end{eqnarray}
 The algorithms we have used to evaluate the self-consistent solutions of the 1RSB Cavity Equations (\ref{eq:r_update},\ref{eq:r_hat_update}) as well as Eq. \eqref{eq:phi_int_x_1} are discussed in the next App. \ref{app:pclock}.  
\section{Population dynamics algorithm for the $p$-clock model applied for the 1RSB Cavity Method with $x^*=1$} 
\label{app:pclock}
Suppose that we have previously run the RSCM algorithm, then we know the distributions of the average messages $\bar \nu$ and $\bar{\hat{\nu}}$. We will indicate these distribution with $\bar P$ and $\bar Q$ respectively. For every value of 
$\tilde{m}= 0, \dots,(p-1)$, we have a population of $N_{\rm pop}$ distributions $\nu$, representing the distribution $R_{\tilde{m}}$, and 
a population of $N_{\rm pop}$ distributions $\hat{\nu}$, representing the distribution $\hat{R}_{\tilde{m}}$. 
In order to update all the $p$ populations of the $\hat{R}_{\tilde{m}}$s we do:
\begin{center}
\hrulefill\\
 \textsc{Population Dynamics, for the 1RSB Cavity Method with $x^* =1$}\\
\hrulefill\\
 \end{center}
1: For $\tilde{m} \in \{0,\dots,p-1\}$\\
2: $\mbox{ }$ For $i \in \{0,\dots,N_{\rm pop}-1\}$\\
3: $\mbox{  }$ Extract $J$ with distribution $\mathcal{P}_{J}$\\
4: $\mbox{  }$ Extract uniform at random $\bar\nu_1,\dots,\bar\nu_{k-1}$ from $\bar P(\nu)$\\
5: $\mbox{}$Given $\tilde{m}$, $J$ and $\bar\nu_1,\dots,\bar\nu_{k-1}$ extract a configuration $\{\tilde{m}_1,\tilde{m}_{k-1}\}$ with probability distribution
  $\pi\bigl(\{\tilde{m}_1,\tilde{m}_{k-1}\}|\tilde{m};J\bigr)$, cf. Eq. \eqref{eq:pi_given_phi}; 
  extract $\nu_1,\dots,\nu_{k-1}$ uniform at random from $R^{(t-1)}_{\tilde{m}_1},\dots,R^{(t-1)}_{\tilde{m}_{k-1}}$.\\
6: $\mbox{  }$ Then, $\hat{\nu}_i$ of the population representing $\hat{R}_{\tilde{m}}^{(t)}$ equals $\hat{f}\left(\nu_{1},\dots,\nu_{k-1}\right)$.\\
7: $\mbox{ }$ End-for\\
8: End-for\\
\hrulefill\\
Then, once we have the $p$ populations $\hat{R}^{(t)}_{\tilde{m}}$, we can update the $p$ populations $R_{\tilde{m}}$ at time $t$ as well:\\
\hrulefill\\
1: For $\tilde{m} \in \{0,\dots,p-1\}$\\
2: $\mbox{ }$ For $i \in \{0,\dots,N_{\rm pop}-1\}$\\
3: $\mbox{  }$ Extract $\hat{\nu}_1,\dots,\hat{\nu}_{c-1}$ uniform at random from the $N_{\rm pop}$ representing $\hat{R}^{(t)}_{\tilde{m}}$\\
4: $\mbox{  }$ $\nu_i$ of the population representing $R^{(t)}_{\tilde{m}}$ equals $f\left(\hat{\nu}_{1},\dots,\hat{\nu}_{c-1}\right)$\\
5: $\mbox{ }$ End-for\\
6: End-for\\
Similar algorithms can be used to evaluate $\phi_{\rm int}$, Eq. \eqref{eq:phi_int_x_1}, and $q_1$, Eq. \eqref{eq:t_q_1}. It is important to remember 
the correct normalization for the $\nu$'s and the $\hat{\nu}$'s now that we are considering the $p$-clock model, rather than the continuous case:
\begin{equation}\label{eq:norm}
\int_0^{2 \pi} d \varphi ~\nu(\varphi) = 1 \quad{ } \rightarrow \quad \sum_{l=0}^{p-1} \nu_l = \frac{p}{2 \pi}  \ .
\end{equation}
$R^{(t)}_{\tilde{m}}$ and $\hat{R}^{(t+1)}_{\tilde{m}}$ are updated until no differences in double precision can be observed in $\Psi$, Eq. \eqref{eq:psi}, and
	$q_1$. To speed up the update algorithm, we have implemented a parallel CUDA-C code scalable also on multi-GPUs ( for $p=32$ we used up to 8 GPUs in parallel). 
\bibliography{Lucabib}
\end{document}